\documentclass[12pt, border=55mm,preprintnumbers]{article} 

\usepackage{graphicx}
\usepackage{url}
\usepackage{multirow}
\usepackage{epsfig}
\usepackage{amsmath}
\usepackage{bm}
\usepackage{times}
\usepackage{graphicx}
\usepackage{color}
\usepackage{slashed}
\usepackage{graphicx}
\usepackage{amsmath}
\usepackage[latin1]{inputenc}
\usepackage{hyperref}
\usepackage{soul}
\usepackage{lipsum}
\usepackage{amssymb}
\usepackage{amsmath}

\usepackage{graphicx}
\usepackage{color} 
\usepackage{relsize} 
\usepackage{lmodern}

\usepackage{hyperref}
\usepackage{enumitem}

\usepackage{float}

\def\bea{\begin{eqnarray}}
\def\eea{\end{eqnarray}}
\def\bean{\begin{equation*}}
\def\eean{\end{equation*}} 
\def\nn{\nonumber}
\def\beaal{\begin{align}}
\def\eeaal{\end{align}}

\begin{document}

\thispagestyle{empty}

\noindent\
\\
\\
\\
\begin{center}
\large \bf Neutron's Dark Secret\,\footnote{Review based on:\\ B.\,Fornal and B.\,Grinstein, \emph{Dark Matter Interpretation of the Neutron Decay Anomaly},  Phys.\,Rev.\,Lett.\,120, 191801 (2018) \cite{Fornal:2018eol} and follow-up publications.}
\end{center}
\hfill
 \vspace*{0.3cm}
\noindent
\begin{center}
{\bf Bartosz Fornal}\\ \vspace{2mm}
{\emph{Department of Physics and Astronomy, University of Utah,\\ 
Salt Lake City, UT 84112, USA}}
\vspace*{0.2cm}
\end{center}
\begin{center}
{\bf Benjam\'{i}n~Grinstein}\\ \vspace{2mm}
{\emph{Department of Physics, University of California, San Diego,\\
9500 Gilman Drive, La Jolla, CA 92093, USA}}
\vspace*{1.5cm}
\end{center}

\begin{abstract}
The existing discrepancy between neutron lifetime measurements in bottle and beam experiments has been interpreted as a sign of the neutron decaying to dark particles.
We summarize the current status\newline of this proposal, including a discussion of 
particle physics models \hspace{6mm} involving such a portal between the Standard Model and a baryonic dark sector.  We also review further theoretical developments around this idea and elaborate on the prospects for verifying the neutron dark decay hypothesis in current and upcoming  experiments.
\end{abstract}

\newpage

\section{Neutron lifetime}

The neutron is one of the most important constituents of matter. It
is absolutely  crucial for the existence of atoms heavier than hydrogen.
Surprisingly, although discovered nearly a century ago \cite{1932RSPSA.136..692C}, the neutron may be hiding a
deep secret related to its decay. The precise value of the free neutron lifetime is an open
question, with the two types of experiments (bottle and beam) providing substantially different answers \cite{2011RvMP...83.1173W,Greene}.

\begin{figure}[t!]
\centerline{\includegraphics[trim={1.8cm 4.6cm 2.5cm 4.5cm},clip,width=9.0cm]{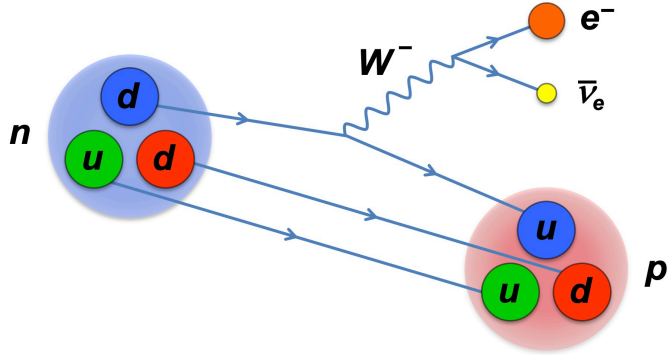}}
\vspace*{0pt}
\caption{\small{Neutron beta decay in the Standard Model.}\protect\label{fig1}}
\vspace{8mm}
\end{figure}

In the commonly accepted description of fundamental interactions, the Standard Model of particle physics \cite{Glashow:1961tr,Weinberg:1967tq,Salam:1968rm,SU(3),Fritzsch:1973pi}, the neutron  predominantly  beta decays to a proton, electron and antineutrino (see, Fig.\,\ref{fig1}). The radiative corrections to this process, involving extra photons in the final state, have a branching fraction ${{\rm Br}(n\to p \,e \, \bar{\nu}_e\gamma)} \approx 10^{-2}$ \cite{Bales:2016iyh}. The  neutron can also decay to a hydrogen atom and antineutrino with ${{\rm Br}(n\to {\rm H}\,\bar{\nu}_e)} \approx 4 \times 10^{-6}$ \cite{Faber:2009ts}. Given that those are the only neutron decay channels within the Standard Model, a proton in the final state is always expected.

A theoretical estimate of the neutron lifetime from beta decays is \cite{Marciano:2005ec,Czarnecki:2018okw}
\bea
\tau_n^{\rm theory} = \frac{4908.6(1.9) \ {\rm s}}{|V_{ud}|^2(1+3\lambda^2)} \ ,
\eea
where $V_{ud}$ is the first element of the Cabibbo-Kobayashi-Maskawa matrix and $\lambda$ is the ratio of the axial-vector to vector current coefficient in the neutron beta decay matrix element. 
Using the average values adopted by the Particle Data Group \cite{PDG2020}: $V_{ud}= 0.97420 \pm 0.00021$ and $\lambda= -1.2724 \pm 0.0023$, the resulting neutron lifetime is $\tau_n = 883.1 \pm 2.7 \ {\rm s}$. A  lattice calculation yields $\lambda = - 1.271 \pm 0.013$ \cite{Chang:2018uxx}, which corresponds to $\tau_n = 885\pm 15\ {\rm s}$.

On the experimental side, there are two types of measurements of the neutron lifetime. In the first method, the  bottle  experiments, ultracold\break neutrons are trapped inside a canister and their number, $N_n$, is determined over time. Given the expected exponential decay pattern,  having recorded the number of neutrons at times $t_i$,
the data are used to extract the bottle neutron lifetime via
\bea
\tau_n^{\rm bottle} = - \frac{N_n}{{dN_n}/{dt}} = \frac{t_i-t_0}{\log\!\left[\!\frac{N_n(t_0)}{N_n(t_i)}\!\right]} \ .
\eea 
In the second method, the beam experiments, cold neutrons are collimated into a beam and the protons from their decays are trapped and counted. Knowing the number of neutrons in the beam, $N_n$, and measuring the rate of their decay to protons, $dN_p/dt$, the beam neutron lifetime is determined using the relation
\bea
\tau_n^{\rm beam} = - \frac{N_n}{{dN_p}/{dt}} \ .
\eea
If the total neutron decay rate, $dN_n/dt$, is equal to the rate of  decay to protons, $dN_p/dt$, then the two lifetimes are equal, $\tau_n^{\rm bottle}=\tau_n^{\rm beam}$. This is the prediction of the Standard Model, in which ${\rm Br}(n\to p +{\rm anything})_{\rm SM} =1$, up to the decay to hydrogen with ${\rm Br}(n\to {\rm H}\,\bar\nu_e) <10^{-5}$.

However, the average neutron lifetime measured in bottle experiments \cite{Serebrov:2004zf,Pichlmaier:2010zz,Steyerl:2012zz,Arzumanov:2015tea,Serebrov:2017bzo,Pattie:2017vsj,Ezhov:2014tna}  is 
\bea
\tau_n^{\rm bottle} = 879.4\pm 0.6 \ {\rm s}\ ,
\eea
whereas the average neutron lifetime from beam experiments \cite{Byrne:1996zz,Nico:2004ie,Yue:2013qrc} is
\bea
\tau_n^{\rm beam} = 888.0 \pm 2.0 \ {\rm s}\ .
\eea
The $\sim 4\,\sigma$ discrepancy between the two types of experiments may be the effect of underestimated, or unaccounted for, systematic errors, but it can also be a sign of new physics. 
The bottle and beam  results can be reconciled if the neutron has  a sizable decay channel with no proton in the final state. In such a scenario,  since the beam and bottle lifetimes are related via
\bea
 \tau_n^{\rm beam} = \frac{\tau_n^{\rm bottle}}{{\rm Br}(n\to p +{\rm anything})}\ ,
\eea
one expects $ \tau_n^{\rm beam}  > \tau_n^{\rm bottle}$, precisely  as the experiments seem to indicate. \newpage

In particular, if the neutron beta decays with a branching fraction of  ${{\rm Br}(n\to p +{\rm anything})} \approx 0.99$, whereas the remaining $1\%$ of the decays are to a final state without a proton, i.e., the neutron undergoes a dark decay with the branching fraction
\bea
{{\rm Br}(n \to {\rm anything}  \ne p)} \approx 0.01 \ ,
\eea
the two experimental results are not in contradiction.  This is the main idea behind the proposal in \cite{Fornal:2018eol}, where it was  shown that phenomenologically viable particle physics models realizing this scenario can be constructed. 
\vspace{5mm}

\section{Neutron dark decay}

In this section we discuss the neutron dark decay from a model-independent perspective. Let us consider a neutron decaying to two or more particles, with at least one of them being a particle beyond the Standard Model, and let us denote by $M_f$ the sum of  masses of the particles  in the final state $f$. For such a decay to take place, one obviously requires $M_f < m_n$. There is also a lower bound on $M_f$ which arises from  forbidding neutron dark decays  in stable nuclei.

\subsection{Nuclear stability}\label{nucls}

Consider a nucleus with atomic and mass numbers $(Z,A)$. A dark decay of one of its neutrons would lead to $(Z,A) \to (Z,A\!-\!1)^* + f$. The excited daughter  nucleus $(Z,A\!-\!1)^*$  would subsequently  de-excite to the ground state by emitting secondary particles, e.g., photons. Experiments like the Sudbury Neutrino Observatory (SNO) \cite{Ahmed:2003sy} and the Kamioka Liquid Scintillator Antineutrino Detector (KamLAND) \cite{Araki:2005jt} conducted searches precisely for such signals and placed a stringent  bound on neutron invisible decays of $\tau(n\to {\rm invisible}) \gtrsim 6\times 10^{29} \ {\rm years}$. A neutron dark decay with a branching fraction $1\%$ occurring inside the nucleus $(Z,A)$ would obviously violate this constraint. 

Nevertheless, if the final state mass $M_f$ is close enough to the neutron mass, i.e., $m_n-S_n< M_f<m_n$, where $S_n$ is the neutron separation energy in the nucleus $(Z,A)$, the nuclear dark decay $(Z,A) \to (Z,A-1)^* + f$ is energetically forbidden, whereas the neutron dark decay $n\to f$ is allowed. Among all stable nuclei, the lowest value of $S_n = 1.664 \ {\rm MeV}$ is observed for $^9{\rm Be}$. This leads to the requirement $M_f > 937.900 \ {\rm MeV}$. A slightly stronger constraint arises from the fact that the excited $^8{\rm Be}^*$ nucleus resulting from the $^9{\rm Be}$ dark decay would quickly decay to two alpha particles, thus lowering the overall energy threshold for the nuclear dark decay by $93 \ {\rm keV}$ \cite{Pfutzner:2018ieu}. Ultimately, this leads to the following constraint on the final state mass of the neutron dark decay,
\bea\label{range1}
937.993 \ {\rm MeV} < M_f < 939.565 \ {\rm MeV} \ .
\eea
This condition assures not only the stability of stable nuclei, but also the stability of the proton with respect to dark decays, which requires $M_f > m_p - m_e = 937.761 \ {\rm MeV}$. Nevertheless, depending on the value of $M_f$, some unstable nuclei with a neutron separation energy lower than that of $^9{\rm Be}$ might undergo dark decays. This will be discussed in Sec.\,\ref{ndd} along with the related experimental searches.

\subsection{Dark decay channels}

The observation that nuclear stability is preserved if $M_f$ falls within the range specified in Eq.\,(\ref{range1}),  opens the door to a new class of models involving various  neutron dark decay channels. Those channels include at least one particle beyond the Standard Model in the final state. Denoting by $\chi$ such a new dark fermion and by $\phi$ a dark scalar or vector, the possible neutron dark decay channels are,
\bea
n \to \chi \, \gamma \ , \ \ \ \ n \to \chi\,\phi \ , \ \ \ \ n \to \chi\,e^+e^- \ , \ \ \ \ ... \ \ \ ,
\eea
where the final states denoted by the ellipses may involve additional dark particles, photons and neutrinos.

From an effective theory point of view, the neutron dark decay is triggered by the Lagrangian terms that  mix the neutron with the dark fermion $\chi$, 
\bea\label{Lagg}
\mathcal{L}^{\rm eff} \supset \varepsilon \,(\bar{n}\,\chi+\bar\chi \,n) \ ,
\eea
where $\varepsilon$ is a model-dependent mixing parameter with mass dimension one. The dark fermion in Eq.\,(\ref{Lagg}) can either appear in the final state of a neutron\newline dark decay, or it can be an intermediate particle in this decay. In the discussion below, we focus on the cases $n \to \chi \, \gamma$ and  $n \to \chi\,\phi$.

\subsection{Neutron $\,\to\,$ dark particle $+$ photon}

The minimal scenario for the neutron dark decay involves one new fermion $\chi$  and a photon in the final state,  i.e., $n \to \chi \, \gamma$. 
Given the condition in Eq.\,(\ref{range1}), the allowed mass range for the particle $\chi$ is
\bea\label{lowerl}
937.993 \ {\rm MeV} < m_\chi < 939.565 \ {\rm MeV} \ .
\eea
Since this is a two-body decay, the photon in the final state is monochromatic and has an energy within the range
\bea
0 < E_\gamma < 1.572 \ {\rm MeV} 
\eea
that depends on the value of $m_\chi$. As the dark fermion mass $m_\chi \to m_n$, the energy of the  photon  $E_\gamma \to 0$.

If the dark fermion $\chi$ is stable, it can be a dark matter particle candidate.\break In such a scenario, to prevent $\chi$ from beta decaying, one requires the condition  $m_\chi< m_p+m_e  = 938.783 \ {\rm MeV}$ to hold. Thus, if $\chi$ is a component of dark matter, the energy of the monochromatic photon coming from the neutron dark decay $n \to \chi \, \gamma$ is expected to be within a more narrow range $0.782 \ {\rm MeV} < E_\gamma < 1.572 \ {\rm MeV}$. 
\vspace{1mm}

At the effective Lagrangian level, the dark decay channel $n \to \chi \, \gamma$ is realized if Eq.\,(\ref{Lagg}) is augmented by the neutron magnetic moment interaction, i.e., the Lagrangian takes the form
 \bea\label{lageff1133}
\mathcal{L}_1^{\rm eff} =\bar{n}\,\Big(i\slashed\partial-m_n +\frac{g_ne}{8 m_n}\sigma^{\,\mu\nu}F_{\mu\nu}\Big) \,n
+  \bar{\chi}\,\big(i\slashed\partial-m_\chi\big) \,\chi + \varepsilon \left(\bar{n}\,\chi + \bar{\chi}\,n\right) \, , \ 
\eea
where $g_n$ is the neutron $g$-factor. In this scenario, if energetically allowed, the neutron undergoes the decay $n \to \chi \, \gamma$  with the rate
\bea\label{rate11}
\Delta\Gamma_{n\rightarrow \chi\gamma} = \frac{g_n^2e^2}{128\pi}\bigg(1-\frac{m_\chi^2}{m_n^2}\bigg)^3  \frac{m_n\,\varepsilon^2}{(m_n-m_\chi)^2}  \ .
\eea
The neutron lifetime discrepancy is resolved  if this decay has a branching fraction $\sim 1\%$, i.e., $\Delta\Gamma_{n\rightarrow \chi\gamma}\sim \Gamma_n/100$, where $\Gamma_n$ is the neutron beta decay rate. A viable particle physics model for this scenario (Model 1) is presented in Sec.\,\ref{model1}.

\newpage

\subsection{Neutron $\,\to\,$ two dark particles}

The pure dark decay channel for the neutron involves two dark particles in the final state: a dark fermion $\chi$ and a dark scalar (or vector) $\phi$. Such a decay $n\to \chi\,\phi$ happens, e.g., through an intermediate dark fermion $\tilde\chi$ which mixes with the neutron and  couples to both $\chi$ and $\phi$. The condition in Eq.\,(\ref{range1}) implies that the final state mass satisfies
\bea\label{15}
937.993 \ {\rm MeV} < m_\chi + m_\phi< 939.565 \ {\rm MeV} \ .
\eea
The lower bound in Eq.\,(\ref{15}) also applies to the intermediate particle ${\tilde\chi}$, that is $m_{\tilde\chi} > 937.993 \ {\rm MeV}$, as otherwise $^9{\rm Be}$ would not be stable with  respect to the neutron dark decay $n\to \tilde\chi\,\gamma$.  In the general case, there are no further constraints on the masses of $\chi$ and $\phi$, i.e., they can both be similar,  \hspace{10mm} $m_\chi \sim m_\phi \sim 469 \ {\rm MeV}$, or there can be a vast hierarchy between them, $m_\chi \gg m_\phi$ or $m_\chi \ll m_\phi$. However, for
the particles  $\chi$ and $\phi$ to be   dark matter candidates,  their stability requires $|m_\chi-m_\phi|< m_p+m_e = 938.783 \ {\rm MeV}$.
\vspace{-4mm}

The effective Lagrangian in this case is given by
 \bea\label{lageff1122}
\mathcal{L}_2^{\rm eff} &=&\bar{n}\,\Big(i\slashed\partial-m_n +\frac{g_ne}{8 m_n}\sigma^{\,\mu\nu}F_{\mu\nu}\Big) \,n
+  \bar{\tilde\chi}\,\big(i\slashed\partial-m_{\tilde\chi}\big) \,\tilde\chi + \varepsilon \left(\bar{n}\,\tilde\chi + \bar{\tilde\chi}\,n\right) \nn\\
&+&  \bar{\chi}\,\big(i\slashed\partial-m_\chi\big) \,\chi \ + \partial_\mu \phi^*\partial^\mu \phi - m_\phi^2|\phi|^2 +(\lambda_\phi \,\bar{\tilde\chi}\,\chi \,\phi + {\rm h.c.}) \ .
\eea
This leads to the neutron dark decay channel $n\to \chi\,\phi\,$ with a rate
\bea\label{rate22}
\Delta\Gamma_{n\rightarrow \chi\phi} = \frac{|\lambda_\phi|^2}{16\pi} \,\sqrt{\left[\big(1-x\big)^2-y^2\right]\left[\big(1+x\big)^2-y^2\right]^3}\, \frac{m_n\,\varepsilon^2}{(m_n-m_{\tilde\chi})^2}  \ ,
\eea
where $x= m_\chi/m_n$ and $y=m_\phi/m_n$. This is the only available dark decay channel if $m_{\tilde\chi} > m_n$. The  neutron lifetime discrepancy is resolved if  the rate $\Delta\Gamma_{n\rightarrow \chi\phi}\sim \Gamma_n/100$. 

If, on the other hand, $m_n>m_{\tilde\chi} > 937.993 \ {\rm MeV}$, the decay channel $n\to \tilde\chi\,\gamma$ also becomes available. The corresponding rate is given by
\bea\label{rate33}
\Delta\Gamma_{n\rightarrow \tilde\chi\gamma} = \frac{g_n^2e^2}{128\pi}\bigg(1-\frac{m_{\tilde\chi}^2}{m_n^2}\bigg)^3  \frac{m_n\,\varepsilon^2}{(m_n-m_{\tilde\chi})^2}  
\eea
and the neutron lifetime puzzle is resolved if $\Delta\Gamma_{n\rightarrow \tilde\chi\gamma}+\Delta\Gamma_{n\rightarrow \chi\phi}\sim \Gamma_n/100$.
A particle physics model for this case (Model 2) is discussed in Sec.\,\ref{model2}.
\vspace{1mm}

Note that a more minimal version of this scenario is obtained by assuming that the intermediate particle $\tilde\chi$ is actually the $\chi$ from  the final state.

\section{Particle physics models}\label{ppmodels}

Here we present the simplest particle physics realizations of the neutron dark decay idea. These models were originally proposed in \cite{Fornal:2018eol}, however, they need to be augmented with extra content to agree with the observed neutron star masses. It is worthwhile to note that such extensions of the Standard Model offer a consistent description of dark matter and can largely explain also the matter-antimatter asymmetry in the universe. 
\vspace{-2mm}

\subsection{Model 1}\label{model1}

The minimal model for the neutron dark decay channel $n\to \chi\,\gamma$ involves only two new particles: $\chi$ and $\Phi$. The Dirac fermion $\chi$ in the final state is a Standard Model singlet. The scalar $\Phi = (3,1)_{-1/3}$ (color triplet, weak singlet, hypercharge $-1/3$) mediates the mixing between $\chi$ and the neutron. The relevant Lagrangian interaction terms are
\bea\label{ll1}
\mathcal{L}_1 \ \supset\  \lambda_q \epsilon^{ijk}\overline{u^c_{L}}_i d_{Rj}\Phi_k+ \lambda_\chi \Phi^{*i}\overline{\chi}\,d_{Ri} \ + \ {\rm h.c.} \ ,
\eea
where $u_R$ and $d_R$ are the right-handed up quark and down quark, respectively, and $u^c_L$ is the charge conjugate of $u_R$.
Baryon number is conserved upon assigning  $B_\chi=1$ and $B_\Phi=-2/3$.

\begin{figure}[t!]
\centerline{\includegraphics[trim={0.0cm 1.0cm 0.0cm 1.15cm},clip,width=8.0cm]{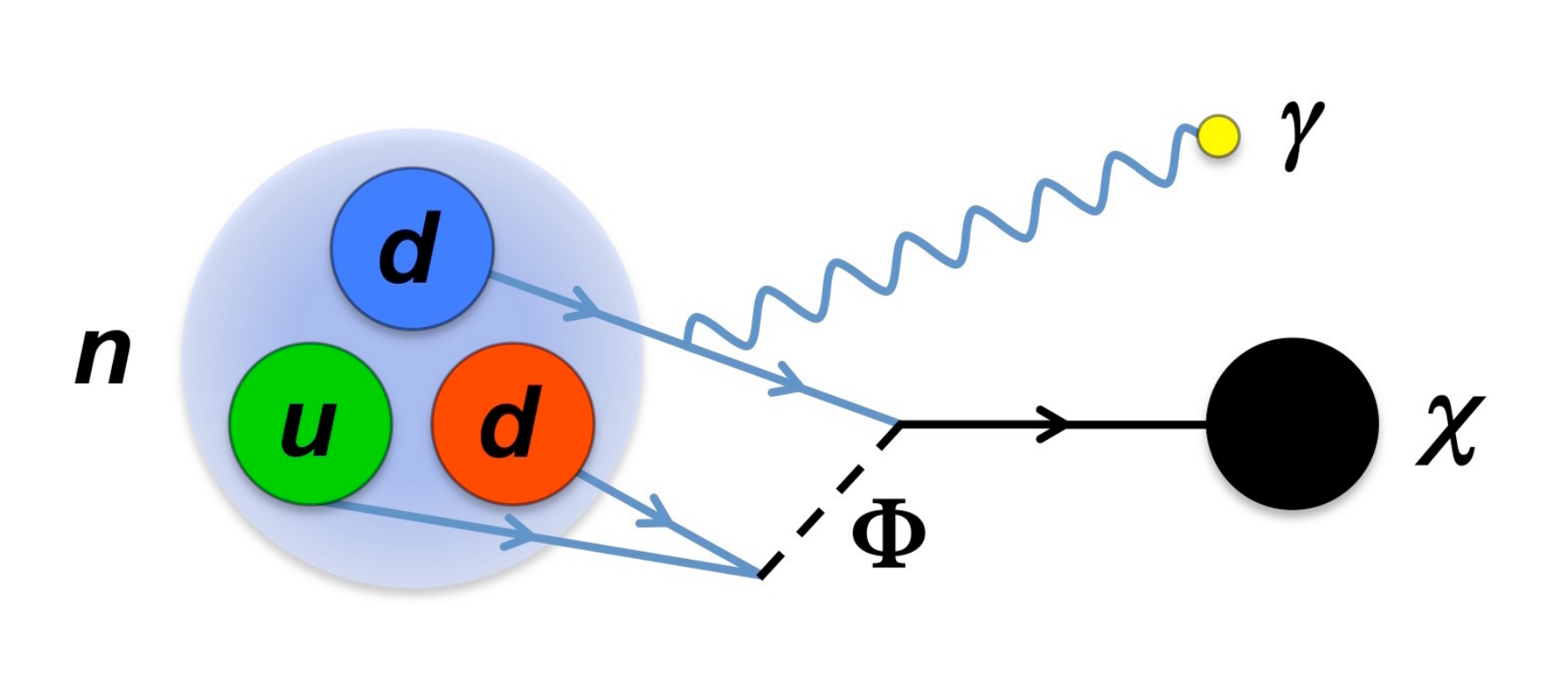}}
\vspace*{6pt}
\caption{\small{Neutron dark decay in Model 1.}\protect\label{fig2}}
\vspace{6mm}
\end{figure}

The neutron dark decay $n\to \chi\,\gamma$ proceeds as shown in Fig.\,\ref{fig2}. The corresponding rate, determined by matching the Lagrangians in Eqs.\,(\ref{lageff1133}) and (\ref{ll1}), is given by the formula in
 Eq.\,(\ref{rate11}) with the mixing parameter
\bea
\varepsilon = \frac{\beta\,\lambda_q\lambda_\chi}{M_\Phi^2}\ ,
\eea
where $\beta = 0.014 \ {\rm GeV}^3$ is calculated from the lattice \cite{Aoki:2017puj}. A neutron dark decay rate of $\Delta \Gamma_{n\to \chi\gamma} \approx \Gamma_n/100$ is obtained for phenomenologically viable choices of parameters. Assuming $m_\chi$ to be at the lower end of the allowed mass range in Eq.\,(\ref{lowerl}), the couplings $\lambda_q$, $\lambda_\chi$ and the mass $M_\Phi$ have to satisfy
\bea\label{b1}
\frac{M_\Phi}{\sqrt{|\lambda_q\lambda_\chi|}} \approx 200 \ {\rm TeV} \ .
\eea 
As long as $M_\Phi \gtrsim 1 \ {\rm TeV}$, collider bounds from $\Phi$ production are preserved. Since $\chi$ is a Dirac fermion carrying baryon number, constraints from dinucleon decays \cite{Gustafson:2015qyo} and neutron-antineutron oscillation \cite{Abe:2011ky} are also avoided.

\subsection{Model 2}\label{model2}

The particle model for the neutron pure dark decay channel, i.e., with no Standard Model particles in the final state, involves four new fields: $\chi$, $\phi$, $\Phi$ and $\tilde\chi$. The final state Dirac fermion $\chi$ and scalar $\phi$, and the intermediate Dirac fermion $\tilde\chi$  are all Standard Model singlets. The scalar $\Phi$ is the same as in Model 1 and mediates  the mixing between $\tilde\chi$ and the neutron. The Lagrangian interaction terms are given by
\bea\label{lag2g}
\mathcal{L}_2 \ \supset \ \lambda_q \epsilon^{ijk}\overline{u^c_{L}}_i d_{Rj}\Phi_k+ \lambda_{\tilde\chi} \Phi^{*i}\overline{\tilde\chi}\,d_{Ri} 
+ \lambda_\phi\overline{\tilde\chi}\,\chi\,\phi \ + \ {\rm h.c.}  \ .
\eea
Choosing $B_\phi=0$ and $B_{\tilde{\chi}}=B_\chi = 1$, baryon number is again conserved.

\begin{figure}[t!]
\centerline{\includegraphics[trim={0.0cm 1.35cm 0.0cm 1.2cm},clip,width=9.0cm]{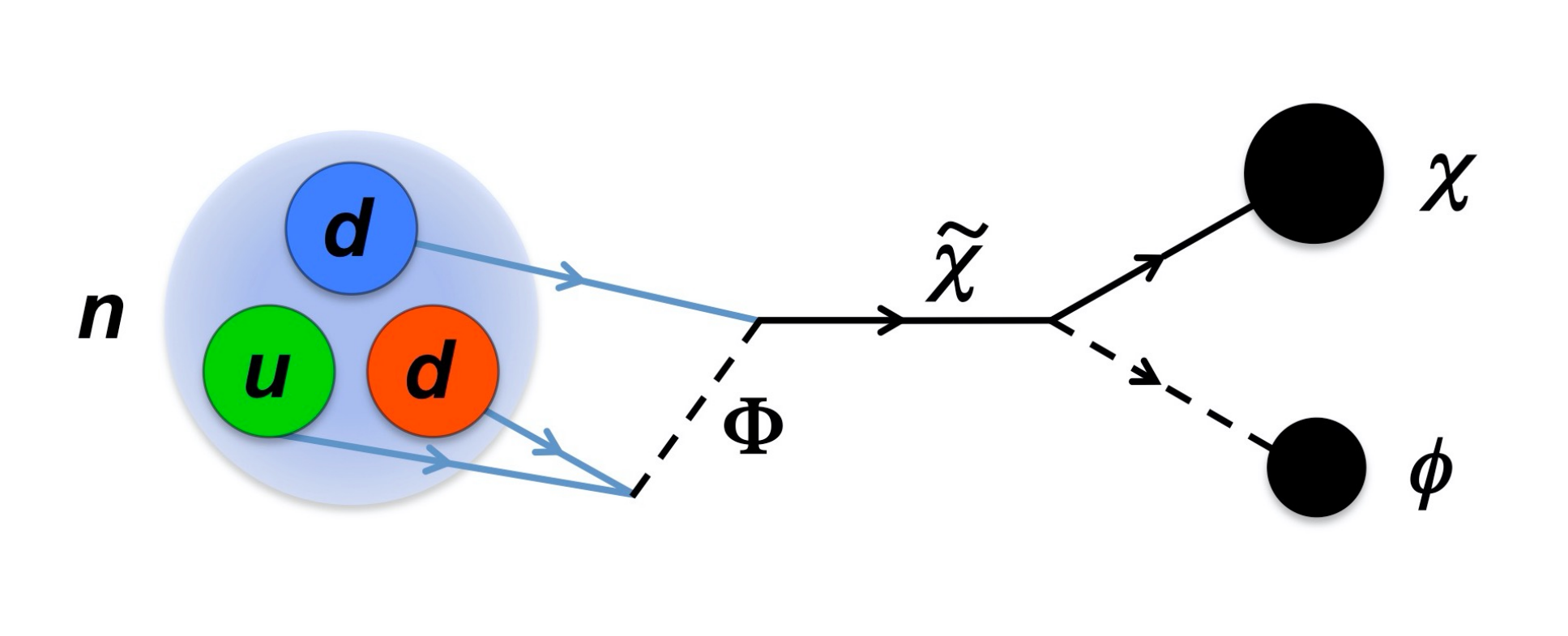}}
\vspace*{8pt}
\caption{\small{Neutron dark decay in Model 2.}\protect\label{fig3}}
\vspace{4mm}
\end{figure}

The neutron dark decay $n\to \chi\,\phi$  is shown in Fig.\,\ref{fig3}. Upon matching the Lagrangians in Eqs.\,(\ref{lageff1122}) and (\ref{lag2g}), the rate for neutron dark decay is given by Eq.\,(\ref{rate22}) with $\varepsilon = {\beta\,\lambda_q\lambda_{\tilde\chi}}/{M_\Phi^2}$. Assuming $m_\chi = 938 \ {\rm MeV}$, $m_\phi \ll m_\chi$ and $m_{\tilde\chi} = 2 \,m_n$, the rate $\Delta \Gamma_{n\to \chi\phi} \approx \Gamma_n/100$ is obtained for
\bea\label{b2}
\frac{M_\Phi}{\sqrt{|\lambda_q\lambda_{\tilde\chi}\lambda_\phi|}} \approx 300 \ {\rm TeV} \ ,
\eea 
which is again consistent with all experiments. This model can be made even more minimal by assuming $\tilde\chi=\chi$.

\section{Experimental developments}

Searches for neutron dark decay began immediately after the idea in \cite{Fornal:2018eol} was proposed. Experimental efforts included looking for direct signatures of the neutron dark decays $n\to\chi\,\gamma$ and $n\to \chi\,e^+e^-$, as well as indirect signals from $n\to \chi\,\phi$ in nuclear decays. Apart from those investigations, a number of proposals have been put forward  regarding complementary search strategies   in current and upcoming experiments. 

\subsection{Photon from $n\to \chi\,\gamma$}

Within one month after the proposal in \cite{Fornal:2018eol} was announced, a dedicated search for the photon signature from the neutron dark decay $n\to\chi\,\gamma$ was conducted at the Los Alamos Ultracold Neutron (UCN) facility 
\cite{Tang:2018eln}. The experiment explored the photon energy range $0.782 \ {\rm MeV} < E_\gamma < 1.664 \ {\rm MeV}$, which corresponds to the case when $\chi$ is a candidate for dark matter. A branching fraction at the level of ${\rm Br}(n\to \chi\,\gamma) = 1\%$ was excluded with an overall significance of $2.2\,\sigma$.

\begin{figure}[t!]
\centerline{\includegraphics[trim={2.5cm 0.5cm 2.5cm 0.0cm},clip,width=9.0cm]{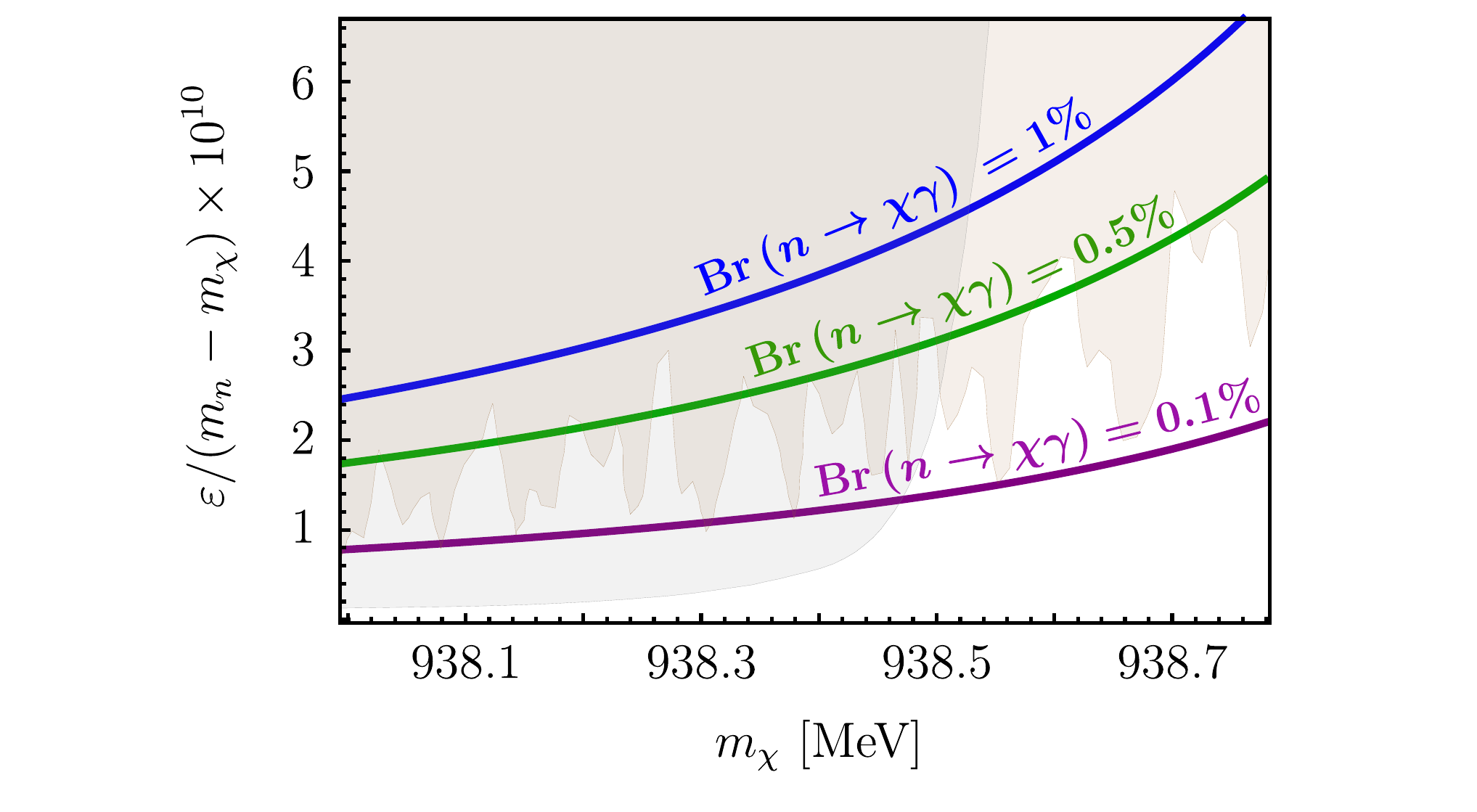}}
\vspace*{5pt}
\caption{\small{The parameter space for the mixing $\varepsilon/(m_n-m_\chi)$ vs. dark matter mass $m_\chi$ excluded at the $90\%$ confidence level by the Los Alamos UCN experiment \cite{Tang:2018eln} (brown-shaded region) and by the Borexino experiment \cite{Agostini:2015oze} (gray-shaded region), as derived in \cite{McKeen:2020zni}. The curves corresponding to the $n\to\chi\,\gamma$ branching fractions of $1\%$, $0.5\%$ and $0.1\%$ are plotted in blue, green and purple, respectively.}\protect\label{fig4}}
\vspace{7mm}
\end{figure}

A more detailed analysis of the UCN data was performed in \cite{McKeen:2020zni}, where the constraints were determined as a function of the dark matter mass, as shown in Fig.\,\ref{fig4}. The brown-shaded region indicates the parameter space excluded by the Los Alamos experiment at the $90\%$ confidence level, whereas the blue line corresponds to the required branching fraction  $1\%$. We note that branching fractions ${\rm Br}(n\to \chi\,\gamma) \lesssim 0.1\%$ are not constrained by the Los Alamos data for any dark matter mass. Also, 
the masses $m_\chi > 938.783 \ {\rm MeV}$, corresponding to $E_\gamma< 0.782 \ {\rm MeV}$, remain to be explored.

\subsection{Electron-positron pair from $n\to \chi\,e^+e^-$}

A search for the electron-positron pairs from the possible neutron dark decay channel  $n\to \chi\,e^+e^-$ was carried out using the Los Alamos UCN data soon afterwards \cite{Sun:2018yaw}, leading to an exclusion of ${\rm Br}(n\to \chi\,e^+e^-) = 1\%$ for energies $E_{e^+e^-} \gtrsim 2\,m_e+100 \ {\rm keV}$. This bound was later improved by the PERKEO experiment \cite{Klopf:2019afh}, extending it to the region $E_{e^+e^-} \gtrsim 2\,m_e+30 \ {\rm keV}$ with a  high confidence level.

\subsection{Nuclear dark decay}\label{ndd}

As discussed in Sec.\,\ref{nucls}, the final state mass of neutron dark decay has to satisfy $M_f > 937.993 \ {\rm MeV}$ in order for stable nuclei to remain stable with respect to  dark decays. The most stringent constraint arises from the stability of $^9{\rm Be}$, whose neutron separation energy $S_n(^9{\rm Be})=1.664 \ {\rm MeV}$.  However, there exist many unstable nuclei with smaller neutron separation energies, $S_n < 1.664 \ {\rm MeV}$. Examples include $^{11}{\rm Li}$ (with $S_n(^{11}{\rm Li}) = 0.396 \ {\rm MeV}$) and $^{11}{\rm Be}$ (with $S_n(^{11}{\rm Be}) = 0.502 \ {\rm MeV}$). For those unstable nuclei, the nuclear dark decay is energetically allowed if
\bea
937.993 \ {\rm MeV} < M_f < m_n - S_n \ .
\eea

It was proposed in {\cite{Fornal:2018eol}} to look for such nuclear dark decays in $^{11}{\rm Li}$ nuclei. Those decays would result in
\bea
^{11}{\rm Li} \to \,^{10}{\rm Li}^* + \chi \to \,^9{\rm Li} + n +\chi \ .
\eea
However, as suggested in \cite{Pfutzner:2018ieu}, a better candidate to search for dark decays is the $^{11}{\rm Be}$ nucleus. 

\noindent
{\bf{$\boldsymbol{^{11}{\rm Be}}$} dark decay -- theoretical motivation}
\vspace{2mm}

\noindent
$^{11}{\rm Be}$ has a halo neutron, which  allows to estimate the nuclear dark decay rate without  calculating any nuclear matrix elements. 
The  main $^{11}{\rm Be}$ decay channels have branching fractions ${\rm Br}(^{11}{\rm Be}\to \,^{11}{\rm B} + e+\bar\nu_e)=97.1\%$ and ${\rm Br}(^{11}{\rm Be}\to \,^{7}{\rm Li}+ \,^{4}{\rm He}+ e+\bar\nu_e)=2.9\%$. There is also a theoretical estimate for the beta-delayed proton emission  ${\rm Br}(^{11}{\rm Be}\to \,^{10}{\rm Be} + p) \sim 2 \times 10^{-8}$ \cite{Borge2013}.
\vspace{1mm}

It was pointed out in \cite{Pfutzner:2018ieu}  that in a past experiment the measured number of $^{10}{\rm Be}$ nuclei resulting from $^{11}{\rm Be}$ decays  exceeded  the theoretical expectation  by a factor of $\sim 400$, i.e., ${\rm Br}(^{11}{\rm Be}\to \,^{10}{\rm Be} + {\rm anything}) \sim 8 \times 10^{-6}$  \cite{Riisager:2014gia}. This experiment  was sensitive only to $^{10}{\rm Be}$ nuclei and could not measure other particles, including protons, in the final state. It was speculated in \cite{Riisager:2014gia} that there must exist an unknown near-threshold proton emitting resonance in $^{11}{\rm B}$,  which would enhance the rate for $^{11}{\rm Be} \to \!^{10}{\rm Be} + p$, making it much larger than the standard expectation. 

Another, quite intriguing hypothesis was put forward in \cite{Pfutzner:2018ieu}, where this large number of $^{10}{\rm Be}$ nuclei was attributed to the  $^{11}{\rm Be}$ nuclear dark decay
\bea
^{11}{\rm Be} \to \,^{10}{\rm Be} + \chi + \phi \ ,
\eea
triggered by the dark decay of the halo neutron. It was shown that the branching fraction ${\rm Br}(^{11}{\rm Be}\to \,^{10}{\rm Be} + \chi+\phi) \sim 8 \times 10^{-6}$ can be accommodated within the framework of Model 2 (discussed in Sec.\,\ref{model2}).
It was later verified in \cite{Ejiri:2018dun}  via a more detailed calculation that such a  $^{11}{\rm Be}$ dark decay is indeed  phenomenologically viable within Model 2 as long as $m_{\tilde\chi}> m_n - S_n = 939.064 \ {\rm MeV}$. Therefore, the  remaining question is whether there are protons in the final state of $^{11}{\rm Be}$ decays and how many. 
\vspace{4mm}

\noindent
{\bf{$\boldsymbol{^{11}{\rm Be}}$} dark decay -- experimental searches}
\vspace{2mm}

\noindent
Two experimental collaborations attempted to find the answer --  one group at CERN--ISOLDE \cite{ISOLDE} and the other  at  ISAC--TRIUMF \cite{Ayyad:2019kna}. While the results 
of the CERN experiment have not yet been published, according to the ISAC--TRIUMF measurement  \cite{Ayyad:2019kna} the number of protons produced in $^{11}{\rm Be}$ decays is roughly equal to the number of 
$^{10}{\rm Be}$ nuclei quoted in \cite{Riisager:2014gia}, i.e., ${\rm Br}(^{11}{\rm Be}\to p + {\rm anything}) \sim 8 \times 10^{-6}$.
This would indicate that the anomalously large number of  $^{10}{\rm Be}$ nuclei measured in \cite{Riisager:2014gia} was due to an undiscovered near-threshold resonance in $^{11}{\rm B}$, which led to an enhanced rate for  beta-delayed proton emission in $^{11}{\rm Be}$ with ${\rm Br}(^{11}{\rm Be}\to \,^{10}{\rm Be} + p) \sim 8 \times 10^{-6}$. The existence of such a resonance has  been  supported by a theoretical calculation \cite{Okolowicz:2019ifb}.

 However, a recent reanalysis  of the  \cite{Riisager:2014gia} data provided an upper limit ${\rm Br}(^{11}{\rm Be}\to \,^{10}{\rm Be} + {\rm anything}) \lesssim 2 \times 10^{-6}$ \cite{Riisager:2020glj}, which is in contradiction with the experimental result of \cite{Ayyad:2019kna}. Further investigation is needed to determine the cause of this disagreement.
 \vspace{-2mm}

\subsection{Beam and bottle experiments}

The fate of the neutron lifetime discrepancy largely depends on the results of the two ongoing neutron lifetime beam measurements: 
one at the National Institute of Standards and Technology (NIST) \cite{DEWEY2009189,Hoogerheide:2019yfu} and the other at the Japan Proton Accelerator Research Complex (J-PARC) \cite{Nagakura:2017xmv,Nagakura:2019xul,Hirota:2020mrd}. Those two efforts are complementary, since NIST is sensitive to protons from neutron decays, whereas J-PARC measures  electrons. If the discrepancy between beam and bottle experiments persists, this will provide additional motivation for investigating neutron dark decay models.

An independent path to resolving the discrepancy would be to construct a single experiment measuring  the bottle lifetime and simultaneously counting the protons from neutron beta decays. A possible strategy would be to install a proton detector in bottle experiments \cite{Fornal:2018mhk}. This idea is currently being implemented by the Los Alamos UCN collaboration within the project:  ``UCN measurement of the Proton branching ratio in neutron Beta decay''  (UCNProBe) \cite{Tang}. 
\vspace{-2mm}

\subsection{Further searches}

Yet another idea is to perform a space-based measurement of the neutron lifetime using a neutron spectrometer like the one  on-board the NASA's MESSENGER spacecraft  \cite{Wilson:2020sjk}. 
The device can measure neutrons generated by galactic cosmic ray spallation of planets' surfaces and atmospheres. The data gathered so far do not provide results competitive with bottle or beam measurements, however, arriving at the accuracy level of $\sim 1 \ {\rm s}$ might be  feasible for future space missions \cite{Wilson:2020sjk}.

A complementary method of investigating neutron dark decay models is to search for the heavy colored particle $\Phi$  at colliders. According to Eqs.\,(\ref{b1}) and (\ref{b2}), for couplings ${\mathcal{O}}(0.1)$ the required mass $M_\Phi \sim {\mathcal{O}}(10 \ {\rm TeV})$, making it possible to produce $\Phi$ at the $100 \ {\rm TeV}$ Future Circular Collider. 
The particle $\Phi$ could potentially be discovered already at the Large Hadron Collider, however, the required couplings in Models 1 and 2 would have to satisfy $|\lambda_q\lambda_\chi |\sim10^{-4}$ and $|\lambda_q\lambda_{\tilde\chi}\lambda_\phi| \sim 10^{-4}$, respectively.

Finally, the idea of looking for signatures of neutron dark decay models in neutrino experiments and dark matter direct detection experiments via dark matter capture by nuclei or dark matter-neutron annihilation will be discussed in Secs.\,\ref{capture} and \ref{annn}.
\vspace{2mm}

\section{Theoretical investigations}

The neutron dark decay proposal sparked not only experimental, but also theoretical activity around the subject.
Some of the theoretical follow-ups include investigating the implications of dark decays for neutron stars and building particle physics models overcoming the resulting constraints via self-interactions in the dark sector or  neutron--dark matter repulsive interaction. Several constructed models of this type can account for not only the dark matter  in the universe, but also the observed matter-antimatter asymmetry. Other theoretical developments include proposing novel signatures of neutron dark decay models to be searched  for in existing and upcoming experiments, as well as investigating  possible connections to other anomalies.

\subsection{Neutron stars}

The presence of a neutron dark decay channel does not threaten the existence of neutron stars \emph{per se}, since such decays would be blocked by the degeneracy pressure of the $\chi$ particles, much like neutron beta decays in a neutron star are blocked by the degeneracy pressure of the electrons and protons. 
However, as pointed out in \cite{Baym:2018ljz,McKeen:2018xwc,Motta:2018rxp},  a neutron dark decay channel softens the neutron star's equation of state. In the minimal setup of Models 1 and 2, this implies that neutron star masses  cannot exceed $0.8$ solar masses, clearly too small compared to the observed cases of two-solar-mass neutron stars.

Nevertheless, the equation of state can become stiffer if Models 1 and 2 \hspace{5mm} are supplemented with extra ingredients. In particular, including strong repulsive self-interactions in the dark sector \cite{Baym:2018ljz,McKeen:2018xwc,Motta:2018rxp} or a repulsive  interaction between the neutron and the dark matter \cite{Grinstein:2018ptl}, does allow neutron stars to reach two solar masses.  Interestingly, strong self-interactions  in the dark sector were proposed many years ago to solve the small-scale structure problems of the $\Lambda$CDM model \cite{Spergel:1999mh}.

\subsection{Dark sector self-interactions}

Strong self-interactions in the dark sector can be realized by introducing  a dark vector gauge boson into the models discussed in Sec.\,\ref{ppmodels}.
An example of such an extended model, involving a dark photon $A'$ and the neutron dark decay channel $n\to \chi\,A'$, was proposed in \cite{Cline:2018ami}.  The Lagrangian of Model 1 given by Eq.\,(\ref{ll1}) was augmented by the following terms,
\bea
\mathcal{L}_1 \ \supset \ -\tfrac14 F'_{\mu\nu}F^{\prime {\mu\nu}} - \tfrac\delta2 F_{\mu\nu}F^{\prime {\mu\nu}} \ ,
\eea
and the covariant derivative was extended to $D_\mu \to D_\mu -i g'A'_\mu$. The coupling between $\chi$ and $A'$ is governed by the gauge coupling $g'$ and leads to repulsive interactions between the $\chi$ particles. It was shown that ${\rm Br}(n\to \chi\,A') = 1\%$ is consistent with neutron star constraints and all other astrophysical bounds for a range of  $g'$ and $\delta$ values. The particle $\chi$ can be a dark matter candidate, however, if one insists on thermal dark matter production, it can account for only $10\%$ of the dark matter in the universe.
A slight extension of this model \cite{Bringmann:2018sbs} was shown to provide a successful framework for explaining the matter-antimatter asymmetry of the universe via low-scale baryogenesis. 
\vspace{1mm}

By giving up the assumption of thermal dark matter production, the astrophysical constraints are considerably relaxed. As shown in \cite{Karananas:2018goc}, if one augments Model 2 from Sec.\,\ref{model2}  with a dark $Z_D$ gauge boson mediating self-interactions in the dark sector via the Lagrangian terms
\bea
\mathcal{L}_2 \ \supset \ g' \bar\chi \,\slashed{Z}_{\!D}\chi - i\,g' Z^\mu_D\left(\phi^* \partial_\mu \phi-\phi\,\partial_\mu \phi^*\right) \ ,
\eea
one can accommodate ${\rm Br}(n\to \chi\,\phi) = 1\%$ and remain consistent with all astrophysical constraints for a wide range of parameters.  In this model $\chi$ can account for all of the dark matter in the universe and its self-interactions solve the small-scale structure problems of $\Lambda$CDM.

\subsection{Neutron-dark matter repulsion}

Another  way of overcoming neutron star constraints   is to introduce extra repulsive interactions between the neutron and the dark matter $\chi$ \cite{Grinstein:2018ptl}.  This can be done by extending  the Lagrangian of Model 2 given in Eq.\,(\ref{lag2g}) with 
\bea
\mathcal{L}_2 \ \supset \ \mu \,H^\dagger H\,\phi + g_\chi\,\bar\chi \,\chi\,\phi \ ,
\eea
where $H$ is the  Higgs field.
Those new terms induce an effective interaction $g_n \bar{n}n\phi$ through the Higgs portal.
Similarly to dark matter self-interactions, this modifies the neutron star's equation of state, allowing for two-solar-mass neutron stars to exist in the presence of the neutron decay channel $n\to\chi\,\phi$.

\subsection{Hydrogen stability}\label{hydro}

In Model 1 with $m_\chi < m_p+m_e = 938.783 \ {\rm MeV}$, i.e., when the particle $\chi$ is a dark matter candidate, not only the neutron but also hydrogen undergoes dark decays  \cite{McKeen:2020zni,Berezhiani:2018eds},
\bea\label{hyddec}
{\rm H} \to \chi\, \nu_e \ .
\eea
As pointed out in  \cite{McKeen:2020zni}, the branching fraction for the radiative hydrogen dark decay ${\rm H} \to \chi\, \nu_e\gamma$ is constrained by the Borexino data \cite{Agostini:2015oze}. This  can be translated into a bound on the neutron dark decay channel $n\to \chi\,\gamma$, with the resulting constraint shown in Fig.\,\ref{fig4}. The gray-shaded region is excluded at  a  high confidence level, reducing the range of dark matter masses  allowed  for  ${\rm Br}(n\to\chi\,\gamma) > 0.5\%$ to $m_\chi \gtrsim 938.5 \ {\rm MeV}$.

\subsection{Dark matter capture}\label{capture}

Also in models with the neutron dark decay channel $n \to \chi\,\gamma$, a novel dark matter detection opportunity arises. Since the dark particle $\chi$ carries baryon number $B_\chi=1$, it can be captured by atomic nuclei through its mixing with the neutron  \cite{Fornal:2020bzz}. This is especially interesting  when $\chi$ is the dark matter particle. As the Earth moves through the dark matter halo in the Milky Way, 
$\chi$ can be captured by  a nucleus $(Z,A)$, forming an excited nucleus $(Z,A+1)^*$, which then de-excites to the ground state  by emitting a single photon or a cascade of photons,
\bea
\chi + (Z,A)\,  \to \, (Z,A+1)^* \,\to\, (Z,A+1)+\gamma_c \ .
\eea
The energy of the cascade depends on the dark matter mass via
\bea
E_c = S_n - (m_n-m_\chi) \ ,
\eea
where $S_n$ is the neutron separation energy for the nucleus $(Z,A+1)$. Thus, $E_c$ differs from the  energy of the standard  cascade triggered by neutron capture by $(m_n-m_\chi)$. 
Prospects for detecting dark
matter capture signals in large volume neutrino experiments
 like the Deep Underground Neutrino
Experiment (DUNE) \cite{Abi:2020wmh}
and in
 dark matter direct detection experiments like PandaX \cite{Cao:2014jsa} and XENON1T \cite{Aprile:2015uzo}, were also investigated in \cite{Fornal:2020bzz}, and the discovery reach is quite promising.

\subsection{Neutron-dark matter annihilation}\label{annn}

A complementary scenario to $\chi$ being the dark matter particle  was considered in \cite{Jin:2018moh,Keung:2019wpw}, where it was assumed that $\chi$ in the final state of neutron dark decay is the antiparticle of dark matter. In this case, $\bar\chi$ is the dark matter particle and carries baryon number $B_{\bar\chi}=-1$.
As a result, it can annihilate with nucleons inside nuclei, leading to spectacular signatures,
\bea
&&\bar{\chi} + n \to \gamma + {\rm meson(s)} \ \ \  \ {\rm in \ Model \ 1}\ ,\nn\\
&&\bar{\chi} + n \to \phi + {\rm meson(s)} \  \ \ \ {\rm in \ Model \ 2} \ ,
\eea
characterized by nonstandard kinematics of the final state, vastly different
from the usually considered nucleon decay case. Such signals can be searched for 
 in
various experiments, e.g.,  Super-Kamiokande \cite{Fukuda:2002uc} and DUNE. 
According to the analysis carried out  in  \cite{Jin:2018moh,Keung:2019wpw}, Model 1 with $\chi$ being the antiparticle of dark matter is excluded by the Super-Kamiokande data, whereas Model 2 remains valid for a large range of parameters. We also note that signatures of this type were considered in a more general context  in \cite{Davoudiasl:2010am,Davoudiasl:2011fj}.

\subsection{Further developments}

Apart from Models 1 and 2 discussed in Sec.\,\ref{ppmodels} and their extensions with dark gauge bosons to account for neutron star constraints, there were also several follow-up proposals for other dark decay channels and specific model realizations, as well as alternative explanations for the neutron lifetime discrepancy. We enumerate a few representative references   below.

In \cite{Elahi:2020urr} a model with a non-Abelian dark gauge group ${\rm SU}(2)_D$ was introduced, in which  the neutron undergoes the dark decay $n\to \chi\,W'$ and the thermal freeze-in mechanism for dark matter production  was implemented.
In  \cite{Barducci:2018rlx} an extension of the neutron dark decay idea to other neutral hadrons, e.g., neutral kaons and $B$-mesons, was proposed.

 An explanation of the neutron lifetime discrepancy within the framework of neutron-mirror neutron oscillations was suggested in  \cite{Berezhiani:2018udo}. However, as shown in  \cite{Fornal:2019eiw}, 
 an extreme breaking of the symmetry between the Standard Model and its mirror copy is required to make the model consistent with experiment. 
 In \cite{Tan:2019mrj}  another model  based on the idea of neutron-mirror neutron oscillations was proposed, but with the neutron dark decay mediated via  non-perturbative effects. 

Other theoretical ideas for the neutron lifetime discrepancy involve an increased rate of neutron-mirror neutron oscillations due to the presence of a magnetic field in beam experiments \cite{Berezhiani:2018eds}, the existence of a large Fierz interference term \cite{Ivanov:2018vit} and the quantum Zeno effect  \cite{Giacosa:2019nbz}. An experimental explanation suggests proton 
 losses in the beam method arising from charge exchange collisions between the residual gas molecules and protons stored in the quasi-Penning trap
\cite{Byrne:2019dhj}.

\subsection{Connection to other anomalies}

Recently, an unexpected excess of keV electron recoil events has been observed by the XENON1T experiment \cite{Aprile:2020tmw}. 
One of the proposed  interpretations of this excess is  boosted dark matter interacting with the electrons  in the detector \cite{Kannike:2020agf,Fornal:2020npv}.
In models with two-component dark matter, such boosted particles can arise from the annihilation of the heavier component.
If  annihilation occurs in the Galactic Center or the Milky Way halo,  a flux of dark matter particles is expected to reach Earth. The XENON1T result can be explained if the velocity of those particles  is $\sim 0.1 \,c$.

We  point out that this explanation of the excess can be realized within Model 2 for the neutron lifetime discrepancy (discussed in Sec.\,\ref{model2}). The $\phi$ particles could annihilate through  $\phi\,\phi^*\to \bar\chi\,\chi$ via a $t$-channel exchange of $\tilde\chi$, producing a flux of $\chi$ particles. Upon reaching the detector, the $\chi$ particles could scatter with electrons through a dark mediator and produce the observed excess in the electron recoil energy distribution. To obtain a boost factor consistent with the XENON1T result, the masses of $\phi$ and $\chi$ should be similar, $m_\chi \sim m_\phi \sim 469 \ {\rm MeV}$, with $\phi$ slightly heavier than $\chi$.

Another possible origin of the excess, also  related to the neutron lifetime discrepancy, was proposed in  \cite{McKeen:2020vpf} within the framework of Model 1 (discussed in  Sec.\,\ref{model1}) extended by a neutral dark fermion $\psi$. It is speculated that when hydrogen atoms (primarily in the oceans) undergo the dark decay as in Eq.\,(\ref{hyddec}), the resulting  particles
$\chi$ propagate through the Earth and decay
in the detector to $\psi$ and photons, which are then absorbed and lead to ionization electrons that can produce the observed excess.

In  \cite{TienDu:2020wks} a connection between  the neutron lifetime discrepancy  and another anomaly was explored. It was suggested that the intermediate particle in the neutron dark decay may be the  $17 \ {\rm MeV}$ gauge boson proposed to explain the anomaly in $^8{\rm Be}$ nuclear transitions \cite{Krasznahorkay:2015iga,Feng:2016jff}.

\section{Summary}

The neutron lifetime discrepancy remains an outstanding problem in modern particle and nuclear physics. The ongoing beam and bottle experiments should soon shed more light on its status. If the discrepancy persists, whether it will be at the current level or at a smaller magnitude, further theoretical and experimental efforts will be needed to pinpoint the exact  origin of the disagreement between the two types of measurements.

The  interpretation of the neutron lifetime puzzle in terms of physics \hspace{5mm} beyond the Standard Model,  postulating the existence of  a neutron dark decay channel, turned out to be quite attractive. Simple phenomenologically viable particle physics models for such a  scenario have been  constructed. Some of them contain in their spectrum a dark matter candidate and  can account for the observed matter-antimatter asymmetry of the universe. 
Such models can also be related to other existing anomalies in particle and nuclear physics, increasing the motivation for their further studies. 

The proposed theories explaining the neutron lifetime discrepancy  involve\break GeV-scale  dark particles carrying baryon number that interact with the neutron via a heavy colored scalar. This interaction introduces a portal between the Standard Model and a baryonic dark sector, providing novel opportunities to probe dark matter in various experiments. 
Some of the new  signatures involve photon cascades from dark matter capture by atomic\break nuclei and nonstandard nucleon annihilation signals, relevant for current and future neutrino and dark matter direct detection experiments. 

Finally, we would like to emphasize that even if the discrepancy between the bottle and beam measurements drops below the level of $1\%$, this would remain interesting, since the neutron can undergo dark decays at a lower rate. This still provides a link to the  dark  sector, giving us hope of experimentally verifying the true nature of dark matter.

\section*{Acknowledgments}

B.F. and B.G. are supported in part by the U.S. Department of Energy Grants
No.~${\rm DE}$-${\rm SC0009959}$ and No.~${\rm DE}$-${\rm SC0009919}$, respectively.
\vspace{5mm}

\bibliographystyle{utphys}
\bibliography{refs}

\end{document}